\documentclass[preprint,superscriptaddress,amsmath,amssymb,aps,showkeys,longbibliography]{revtex4-2}
 \pdfoutput=1
\usepackage{graphicx}
\usepackage{dcolumn}
\usepackage{bm}
\usepackage{hyperref}
\usepackage{booktabs}
\hypersetup{
    colorlinks=true,
    linkcolor=blue,    
    citecolor=blue,     
    urlcolor=blue   
}

\usepackage[utf8]{inputenc}
\usepackage[T1]{fontenc}
\usepackage{xfp}

\begin{document}

\preprint{APS/123-QED}

\title{Doping induced magnetism and half-metallicity in nanoribbons of quartic dispersion materials}

\author{Emin Aliyev}
\altaffiliation[Current address: ]{Max Planck Institute for the Structure and Dynamics of Matter, Hamburg 22761, Germany}
\affiliation{%
 UNAM-Institute of Materials Science and Nanotechnology, Bilkent University, Ankara 06800, Turkey\\}
\author{Arash Mobaraki}%
\affiliation{%
 Department of Physics, Aydin Adnan Menderes University, Aydin 09010, Turkey\\}%

\author{Hâldun Sevinçli}
\thanks{Contact author: sevincli@fen.bilkent.edu.tr}
\affiliation{
Department of Physics, Bilkent University, Ankara 06800, Turkey} 
 
\author{Seymur Jahangirov}%
\thanks{Contact author: seymur@unam.bilkent.edu.tr}
\affiliation{%
 UNAM-Institute of Materials Science and Nanotechnology, Bilkent University, Ankara 06800, Turkey\\}


\date{\today}

\begin{abstract}
Two-dimensional (2D) quartic dispersion materials are known to develop magnetization upon doping. Here we conduct a systematic investigation of magnetization in hole-doped quartic dispersion materials (GaS, InSe, TiO$_{2}$), focusing on the effects of structural confinement from 2D monolayers to quasi-one-dimensional nanoribbons (NRs).  Upon hole doping, these NRs develop itinerant magnetization across a broad range of carrier densities and display half-metallic behavior. The spin-polarization energies ($E_{sp}$) of these NRs enhance remarkably relative to their 2D counterparts, with maximum increase being in the case of TiO$_{2}$ from 31 to 103 meV/carrier. The $E_{sp}$ strongly depends on the degree of localization of the magnetic moments along the width of NRs, which is determined by edge passivation and ribbon width. Strong deformation of the topmost valence bands at higher dopings indicates deviation from the Stoner mechanism.
\end{abstract}

\maketitle


\section{\label{sec:level1} Introduction}

Over the last decade, various low-dimensional materials, possessing high spin polarization~\cite{guo2025spin,gurung2024nearly,xie2025emerging,volnianska2010magnetism,meng2022hole}, have been investigated for utilization in the emerging field of spintronics, where charge and electron spin, along with their magnetic moments, are harnessed for energy-efficient information processing and storage~\cite{ahn20202d}. One of the principal methods for achieving new electronic and spintronic properties in non-magnetic materials is charge doping~\cite{cao2015tunable,meng2022hole}. Particularly, the electrostatic doping implemented in field-effect transistor (FET) architectures allows for defect‑free, reversible tuning of doping levels in low‑dimensional materials, offering a pathway toward more stable and durable device operation~\cite{meng2022hole}. 

Numerous theoretical studies have employed density functional theory (DFT) to investigate and predict low-dimensional materials that could be potential candidates for spintronic applications~\cite{cao2015tunable,iordanidou2018hole,gao2018edge,ma2025electrically}. In a pioneering study by Cao et al.~\cite{cao2015tunable}, the electrostatic hole doping in the GaSe 2D monolayer was shown to result in spin splitting and the emergence of itinerant magnetism persisting in a wide range of carrier densities, as well as half metallicity, meaning that only a single spin channel crosses the Fermi level ($E_{f}$), making it suitable for spintronic applications. The underlying reason for such properties is the Mexican-hat-shaped (MHS) topmost valence band, exhibiting quartic dispersion, which results in Van Hove singularity (VHS) in the density of states (DOS) near $E_{f}$ that, in turn, leads to ferromagnetic instabilities~\cite{cao2015tunable,chen2018ferromagnetism, sevinccli2017quartic}. In subsequent studies~\cite{meng2022hole, iordanidou2018hole}, it has been shown that such behavior upon hole doping is also characteristic of other 2D materials with MHS electronic band structures. Due to the sharp singularity in DOS, exchange energy overcomes kinetic energy, resulting in spin splitting as described by the Stoner mechanism~\cite{stoner1939collective}. Notably, the MHS electronic band structure emerges only upon dimensional reduction of these materials from 3D to 2D, below a certain critical number of layers~\cite{rybkovskiy2014transition}. 

This naturally leads one to question the implications of further confinement from 2D to 1D on hole-doped electronic and magnetic properties. For example, while doping does not induce magnetization in 2D graphene, it can do so in graphene NRs owing to the strong band-edge singularity. Employing DFT calculations, it was demonstrated that narrow dihydrogenated graphene zigzag-edge nanoribbons (ZNRs) exhibit robust itinerant magnetism and half-metallicity upon electrostatic hole doping in the carrier density range of about $10^{13}$~cm$^{-2}$, with the energy difference per carrier between doped ferromagnetic and non-magnetic states, also known as spin-polarization energy ($E_{sp}$), reaching 17 meV/carrier ~\cite{gao2018edge}. $E_{sp}$ was found to decrease with increasing width of NRs, as quantum confinement effects diminish. While graphene NRs have been extensively investigated, the emergence of diverse two-dimensional materials has broadened the exploration of NR structures beyond carbon systems. In particular, group III metal monochalcogenide (MXs, M = Ga, In and X = S, Se) 1D MX nanostructures have gained substantial research interest in recent years~\cite{yu2024quasi,gutierrez2024unveiling, li2023emergence, hu2022prediction, wen2019two, bergeron2021polymorphism,grzonka2021novel,li2020various} and a variety of these nanostructures, including NRs, have been fabricated, showing potential for advanced nanophotonic and optoelectronic applications~\cite{sutter2021optoelectronics, hauchecorne2021gallium, sutter2020vapor, wu2020controlled, xiong2016one, shen2009vapor, panda2008synthesis,arora2021recent}. Furthermore, MHS band edges of 2D MX systems make them promising for the question at hand.

In our previous work~\cite{aliyev}, we demonstrated that in MX ZNRs, including InSe and GaS, the 1T phase is thermodynamically more favorable until a critical width, which can reach up to 34 nm. Furthermore, unlike metallic 1H ZNRs, 1T ZNRs remain semiconductors and retain MHS top valence bands. More importantly, hydrogenation energies ($E_{H}$) of 1T InSe NRs are positive (unfavorable), and due to the edge-localized states, the 1T unpassivated ZNRs possess nearly flat top valence bands. These electronic properties present compelling opportunities for exploiting 1T ZNRs in spintronic applications. In addition, to explore these phenomena on materials beyond the group III-VI family, we also investigate TiO$_{2}$ NRs. 2D 1T TiO$_{2}$ was also shown to exhibit MHS upper valence bands, and exceptionally high $E_{sp}$, which is a few-fold higher than in 2D MXs~\cite{meng2022hole}.

Building upon these findings, this study investigates the effect of structural confinement of 2D 1T phase InSe, GaS, and TiO$_{2}$ to 1D ZNRs on electronic and magnetic properties under electrostatic hole doping. Studying the correlations between real-space hole distributions along the width of NRs and corresponding electronic band dispersions, we deduce that in NRs, strong localization of magnetization leads to a rise in $E_{sp}$, relative to its 2D counterpart. We observe more than threefold increase in $E_{sp}$ with respect to their 2D counterparts in InSe and TiO$_{2}$ ZNRs, reaching a value of 103 meV/carrier.

 \begin{figure}[h!]  
  \centering  
  \includegraphics[width=0.55\linewidth]{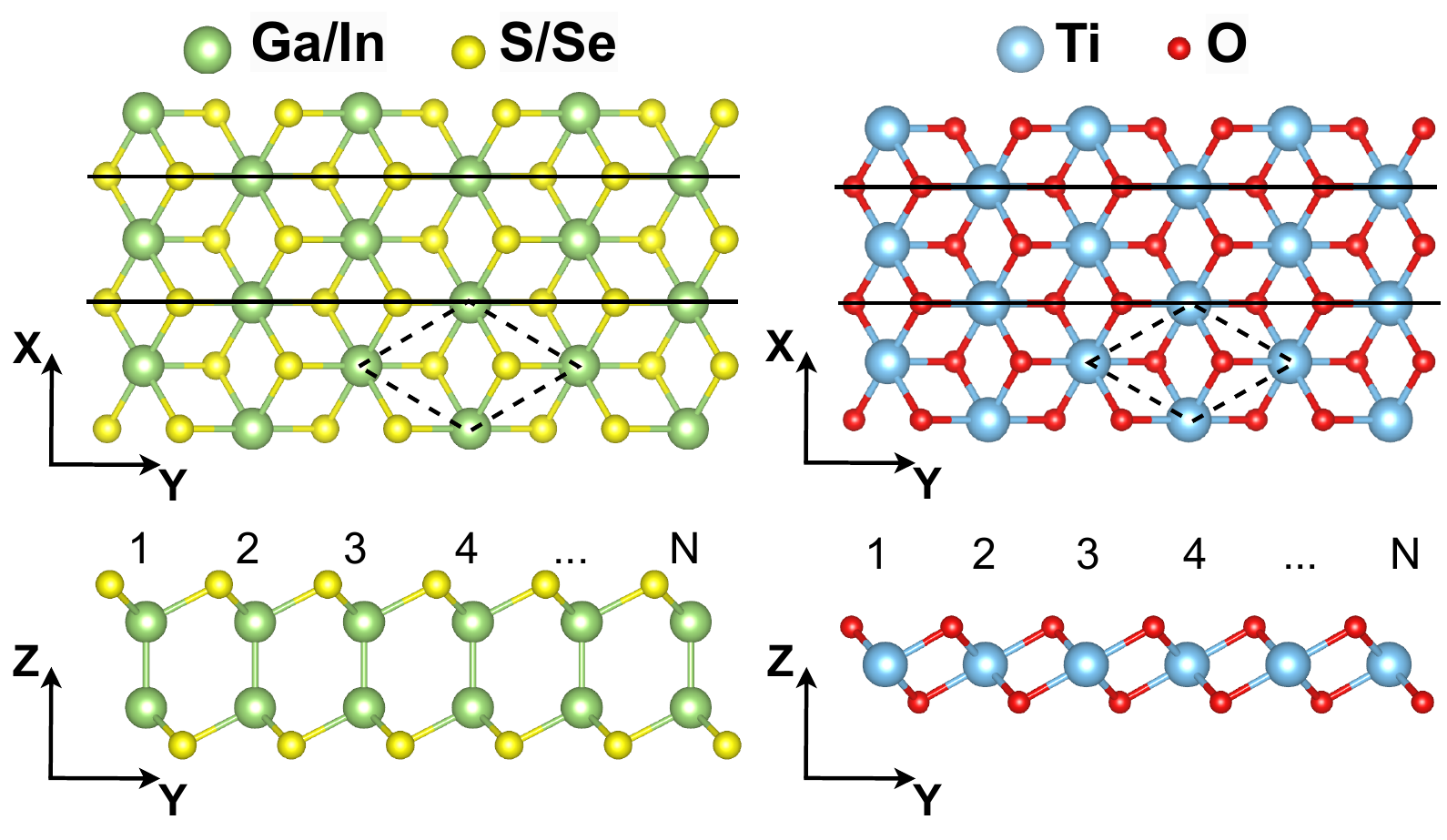}
  \caption{Schematic representations of 1T MX and TiO$_{2}$ ZNRs and the unit cells of the corresponding 2D monolayers.}
  \label{fig:1}
  \end{figure}
  
\section{\label{sec:level2} COMPUTATIONAL DETAILS}

We perform first-principles density functional theory (DFT) calculations using plane-wave basis sets. Projector augmented-wave pseudopotentials~\cite{kresse1999ultrasoft,blochl1994projector} within the generalized gradient approximation of Perdew, Burke, and Ernzerhof~\cite{perdew1996generalized,kresse1993ab,kresse1996efficiency} are used. Spin-orbit coupling (SOC) effect is considered in the case of InSe ~\cite{steiner2016calculation}. To eliminate interactions with periodic images, a vacuum of 15~\text{\AA} is considered in non-periodic directions.  A plane-wave basis set with a 500~eV energy cutoff is employed. Electrostatic hole doping is achieved by removing electrons and compensating with a uniform jellium background of opposite charge to preserve neutrality. The structures are fully relaxed at every carrier density with a $\mathrm{\Gamma}$-centered k-point mesh used for sampling the Brillouin zone (BZ) with a k-spacing of $< 0.18~\text{\AA}^{-1}$. The convergence tolerance for the total energy of the system is set to be less than $10^{-7}$~eV, and the Hellmann-Feynman forces are minimized to be less than $10^{-2}$ eV/\text{\AA}. To obtain well-converged energies and magnetic moments in doped structures, we sample the BZ with dense $\mathrm{\Gamma}$-centered k-point mesh with $< 0.05~\text{\AA}^{-1}$ k-spacing, and apply a Gaussian smearing of 0.001~eV for the electronic occupations. All DFT calculations in this study are conducted using the Vienna Ab initio Software Package (VASP)~\cite{kresse1996efficient}.

\section{RESULTS AND DISCUSSIONS}

The selection of materials is based on the presence of a quartic band dispersion in 2D, which facilitates magnetization upon doping~\cite{cao2015tunable,meng2022hole}, and on the inversion-symmetric 1T phase, which preserves both the band gap and quartic dispersion under confinement to pristine ZNRs~\cite{aliyev} and thus permits doping strategies analogous to those employed in the 2D case to induce magnetization. GaS and InSe in their 1H 2D monolayer form exhibit comparable maximum $E_{sp}$ of about 11 meV per carrier~\cite{meng2022hole,iordanidou2018hole}. The magnetic and electronic properties of these 2D materials show only negligible dependence on the underlying 1T or 1H lattice structure: both phases exhibit virtually identical $E_{sp}$ trends and magnitudes as a function of carrier doping and exhibit half-metallicity and itinerant magnetization under comparable doping conditions. However, ZNRs of InSe and GaS have an important distinction. In that, $E_{H}$ in the case of GaS ZNR is negative, implying the favorability of hydrogen-saturated edges and a consequently more homogeneous charge distribution. On the other hand, a positive $E_{H}$ in InSe implies an unpassivated ground state with an edge-localized highest valence band, which results in a higher increase in $E_{sp}$. To demonstrate the generality of the underlying physical mechanisms of $E_{sp}$ enhancement upon confinement beyond the III-VI family, TiO$_{2}$ is also studied as a reference material. It similarly retains the 1T phase and exhibits MHS top valence band, while demonstrating an $E_{sp}$ nearly three times greater than that of GaS and InSe~\cite{meng2022hole}.  

\begin{figure}[h!]  
  \centering  
  \includegraphics[width=0.55\linewidth]{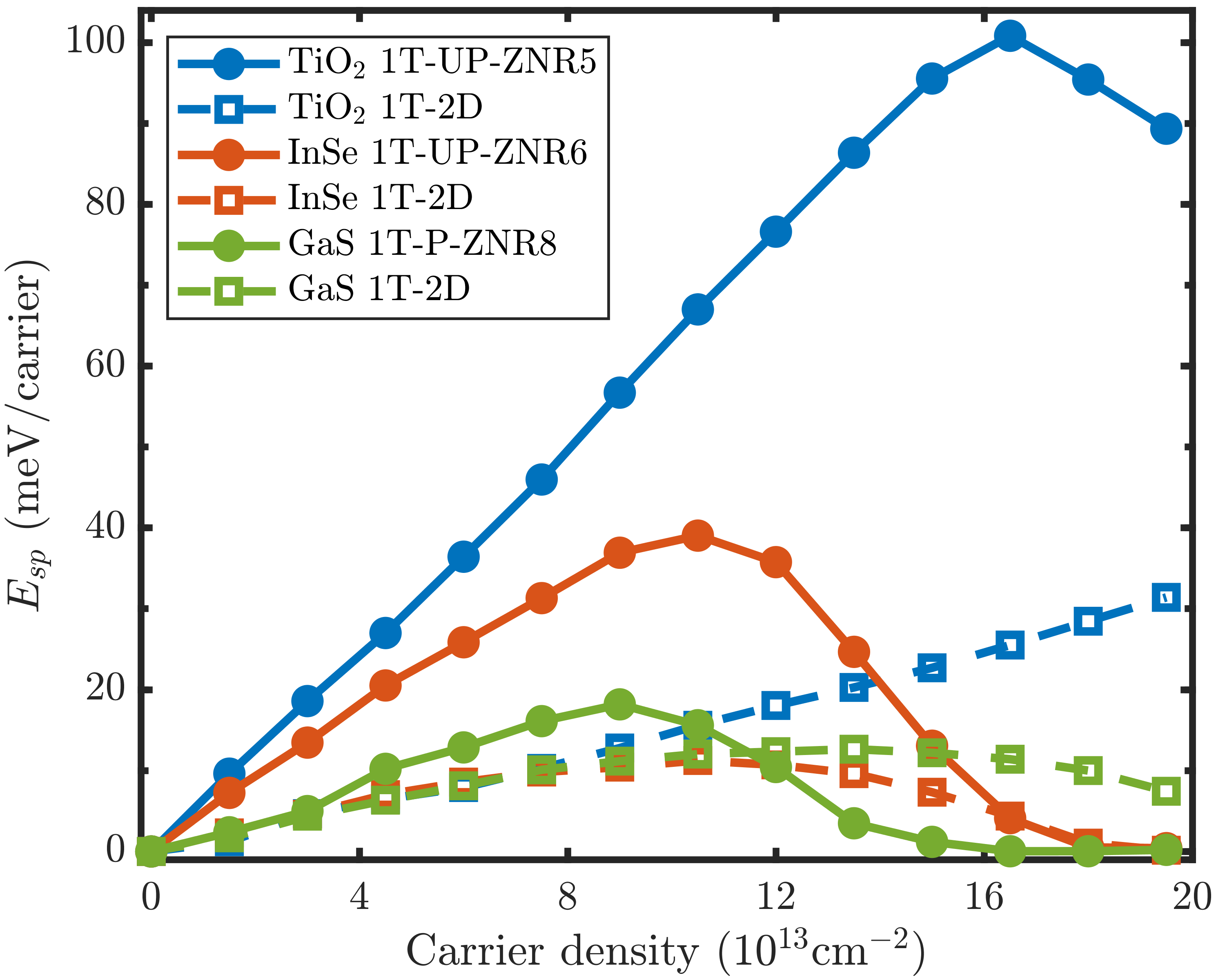}
  \caption{Spin polarization energies of 2D 1T-phase TiO$_{2}$, InSe, and GaS, along with selected ZNR configurations.}
  \label{fig:2}
\end{figure}

The ball and stick representation of the nanoribbon structures is presented in Fig.~\ref{fig:1}. We define \textit{N} as the number of 2D rectangular unit cells forming the NR. Our calculations encompassed ZNRs with widths ranging from $N$ = 5 to 10, and our discussions focus on the ZNR configurations that exhibit the highest $E_{sp}$ enhancements relative to 2D. These main findings are summarized in Fig.~\ref{fig:2}. The highest $E_{sp}$ magnitudes that can be achieved for 2D GaS, InSe, and TiO$_{2}$ in the considered doping range are 12.6, 11.2, and 31.4~meV/carrier, respectively.  We demonstrate that upon structural confinement to 1D ZNRs, these materials exhibit stronger magnetization, which is reflected through an increase in $E_{sp}$ to 17, 39, 103~meV/carrier, respectively.

In the subsequent paragraphs, we will dive deeper into each of these cases. However, the overarching theme of those discussions is the increasing and decreasing profiles of $E_{sp}$ versus carrier density which can be attributed to the position of the Fermi level with respect to van Hove singularities. In that, when the Fermi level starts crossing a singularity upon hole doping,  $E_{sp}$ increases in proportion to the strength of the singularity and decreases when the Fermi level moves towards the tail, away from the singularity. $E_{sp}$ decreases also when the Fermi level starts crossing a singularity of the majority spin. In fact, the same mechanism is already present in 2D. In 1D, both the strength and the number of singularities are increased. Moreover, in contrast to 2D, in 1D the bands and hence the density of states (DOS) profiles are deformed substantially with doping. Also, localized states with high DOS emerge at the edges or at the center of the NRs, depending on the material and the geometry. All these effects create higher and more complicated $E_{sp}$ profiles with respect to doping.

\begin{figure*}[h!]  
  \centering  
  \includegraphics[width=0.75\textwidth]{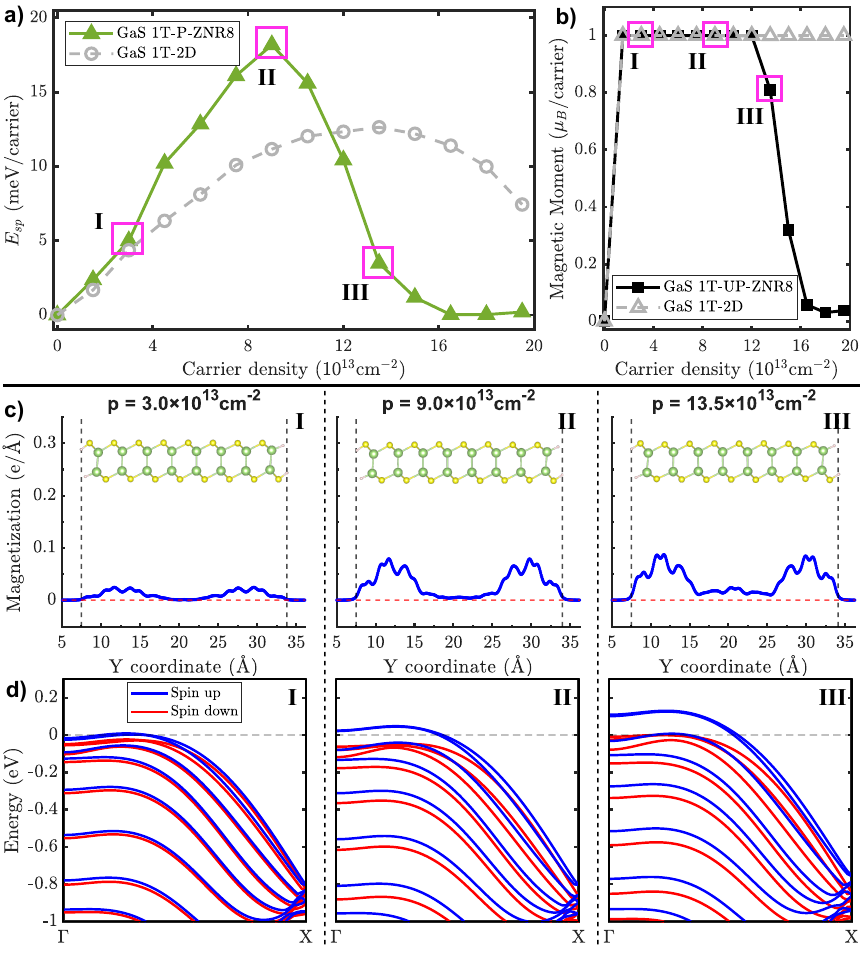}
  \caption{GaS 1T-P-ZNR8 and 1T-2D a) spin polarization energies and b) magnetic moments per carrier as a function of the carrier density. c) magnetization profiles along the width of NR and d) spin-polarized band structures at selected carrier densities.}
  \label{fig:GaS1TPZNR8}
\end{figure*}

As shown in Fig.~\ref{fig:GaS1TPZNR8}(a), upon doping 1T 2D GaS, the $E_{sp}$ value is comparable to that reported for its 1H phase~\cite{meng2022hole}. Upon confinement to 1D ZNR, there is an enhancement in $E_{sp}$ values relative to 2D, already starting from low doping rates. At a carrier density of $p = 4.5 \times 10^{13}$~cm$^{-2}$, we observe a substantial increase in $E_{sp}$ from 6 meV in 2D to 10.5 meV in ZNR, while this NR exhibits maximum $E_{sp}$ at $p = 9 \times 10^{13}$~cm$^{-2}$, where we observe about 50\% enhancement in $E_{sp}$, reaching 17~meV/carrier. In addition, as shown in Fig.~\ref{fig:GaS1TPZNR8}(b), unlike $E_{sp}$, magnetic moments do not exhibit significant dependence on carrier density, as they reach 1~$\mu_{B}$/carrier already at a low doping rate and saturate, showing a decrease only at very high dopings.

To illustrate the origins of the above observations, we depict the magnetization profiles and electronic band structures at selected doping rates in Figs.~\ref{fig:GaS1TPZNR8}(c) and (d), respectively. From Fig.~\ref{fig:GaS1TPZNR8}~(c), it can be inferred that at moderate doping rates, where only the nearly degenerate MHS topmost valence bands cross the $E_F$, the magnetization profiles exhibit broad peaks near the edge of NRs, which intensify with increasing doping rate, while the center remains mainly nonmagnetic. At higher dopings, the Fermi level starts moving away from the singularity which results in decrease of $E_{sp}$ as seen in Fig.~\ref{fig:GaS1TPZNR8}~(d-II) and Fig.~\ref{fig:GaS1TPZNR8}~(d-III).

From Fig.~\ref{fig:GaS1TPZNR8}(d), it is evident that at carrier densities, where the magnetic moments remain saturated at 1~$\mu_{B}$/carrier, the band structure exhibits half-metallicity, when only the minority spin (spin-up) channel crosses the $E_{f}$. On the other hand, at high doping rates, the majority spin (spin-down) channel also crosses the $E_{f}$, and we observe a decrease in magnetic moments that eventually vanish, which is in line with previous reports on hole-doped 2D materials.

\begin{figure*}[h!]  
  \centering  
  \includegraphics[width=0.75\textwidth]{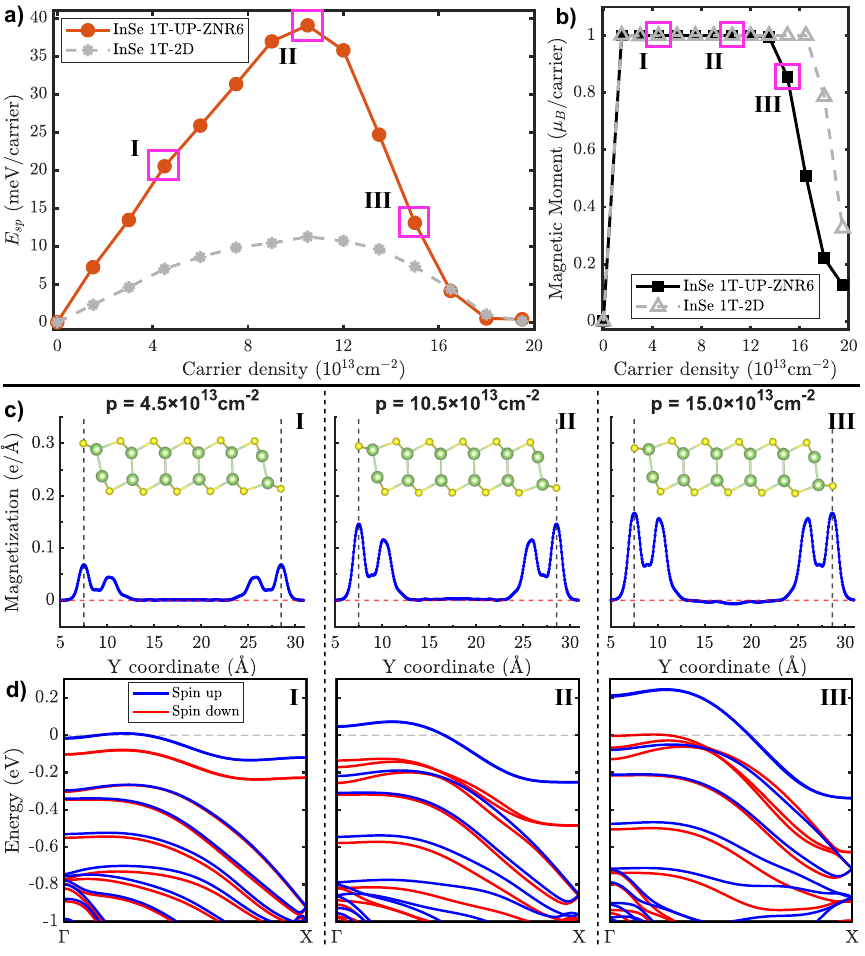}
  \caption{InSe 1T-UP-ZNR6 and 1T-2D a) spin polarization energies and b) magnetic moments per carrier as a function of the carrier density. InSe 1T-UP-ZNR6 c) magnetization profiles along the width of NR and d) spin-polarized band structures at selected carrier densities.}
  \label{fig:InSe1TUPZNR6}
\end{figure*}

To examine the hole doping of III-VI NRs in more detail, we move to the case of InSe, which has comparable $E_{sp}$ to GaS in 2D, but exhibits significantly higher $E_{sp}$ upon confinement to 1D (see Figs.~\ref{fig:2} and~\ref{fig:InSe1TUPZNR6}), reaching 39 meV/carrier at $p = 10.5 \times 10^{13}$~cm$^{-2}$ at $N$ = 6, a fourfold increase relative to 2D. It is worth noting that, as depicted in Fig.~\ref{fig:sup_INSESOC}, inclusion of SOC effects does not substantially alter the observed trends in $E_{sp}$, leading to only a reduction of approximately 10\% in maximum spin-polarization energy ($E^{NR_{max}}_{sp}$) achieved. Notably, unlike GaS 1T-P-ZNRs, InSe 1T-UP-ZNRs exhibit a flatter dispersion of MHS, with nearly degenerate topmost valence bands that stem from edge-localized states due to the absence of hydrogen passivation. Consequently, upon hole doping, the magnetization profile peaks exhibit stronger localization at the edges, as shown in Fig.~\ref{fig:InSe1TUPZNR6}~(c), which intensifies further with increasing carrier density. Since the holes are unevenly distributed across the nanoribbon, the onsite electron-electron interactions are non-uniform in contrast to 2D or a nanotube. This variation could explain the observed deformation at the top valence bands. Increasing the NR width and beyond a certain doping, carriers also appear in the center region of the NR. For instance, in the wider InSe 1T-UP-ZNR10, $E_{sp}$ is higher than in InSe 1T-UP-ZNR6 till carriers appear in the center of NR at $p = 7.5 \times 10^{13}$~cm$^{-2}$ as shown in Fig.~\ref{fig:sup_InSe1TUPZNR10} (c). Moreover, a similar behavior with increasing width is also observed in the case of GaS 1T-P-ZNRs (see Figs.~\ref{fig:GaS1TPZNR8} and~\ref{fig:sup_GaS1TPZNR10}). These observations suggest a correlation between localization and $E_{sp}$. 

\begin{figure*}[h]  
  \centering  
  \includegraphics[width=0.75\textwidth]{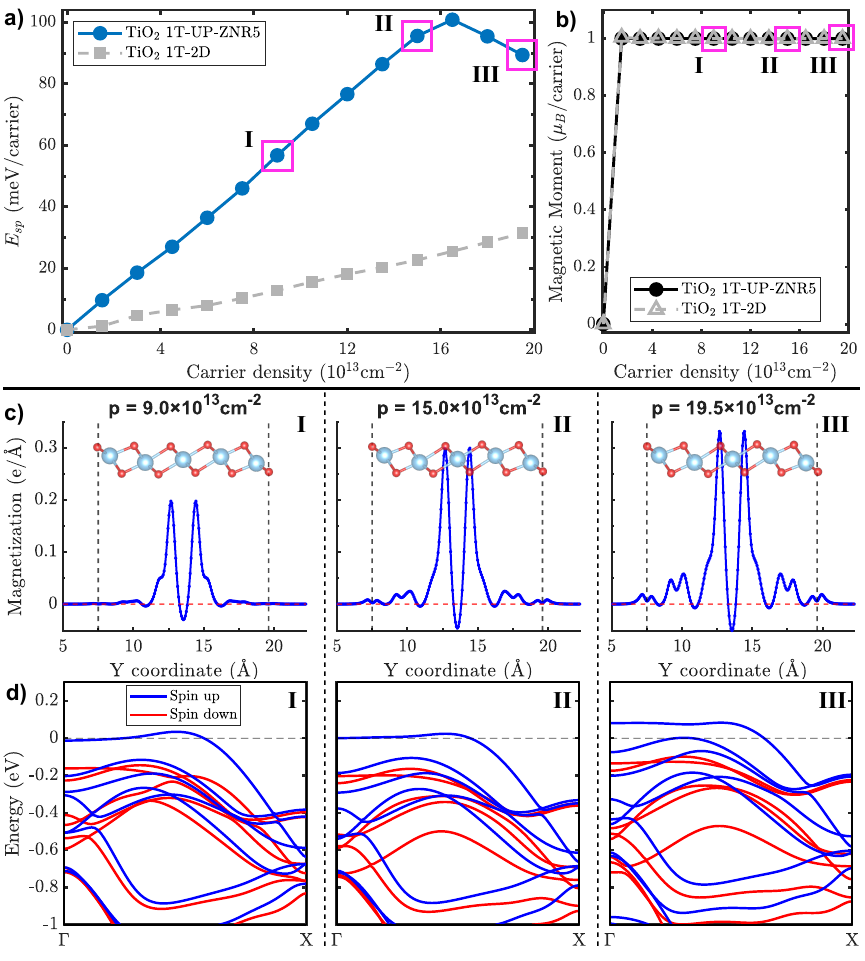}
  \caption{TiO$_{2}$ 1T-UP-ZNR5 and 1T-2D a) spin polarization energies and b) magnetic moments per carrier as a function of the carrier density. TiO$_{2}$ 1T-UP-ZNR5 c) magnetization profiles along the width of NR and d) spin-polarized band structures at selected carrier densities.}
  \label{fig:TiO1TUPZNR5}
\end{figure*}

To further investigate the aforementioned correlation between edge-localization of magnetization profile and $E_{sp}$ beyond the III-VI family, we also studied TiO$_{2}$ 1T-UP-ZNRs (see Fig.~\ref{fig:TiO1TUPZNR5}). It should be noted that these TiO$_{2}$ ZNRs in the narrow width regime were shown to exhibit distinct electronic properties depending on \textit{N} being odd or even, due to interaction of edge O atoms being attractive or repulsive~\cite{he2010first}, which in turn affects their magnetization profiles and $E_{sp}$ vs. carrier density dependence. At moderate carrier densities, magnetization of odd ZNRs displays center localization (see Fig.~\ref{fig:TiO1TUPZNR5}), while in even ZNRs it is relatively uniform (see Fig.~\ref{fig:sup_TiO1TUPZNR6}). It is worth noting that, unlike hole-doped GaS and InSe NRs, which show ferromagnetic behavior, where spins of both metal and chalcogen atoms are aligned along the same direction, in the case of TiO$_{2}$ NRs, upon hole doping, ferrimagnetic behavior is exhibited, with an antiparallel alignment of the spins of Ti and O atoms. TiO$_{2}$ 1T-UP-ZNR5, similar to InSe 1T-UP-ZNR6, exhibits roughly a fourfold increase in SPE relative to 2D, reaching 103 meV/carrier (see Fig.~\ref{fig:TiO1TUPZNR5} (a)) and nearly flat band topmost valence band (see Fig.~\ref{fig:TiO1TUPZNR5} (d)). However, as depicted in (see Fig.~\ref{fig:TiO1TUPZNR5} (c)), unlike InSe ZNRs, the TiO$_{2}$ 1T-UP-ZNR5 exhibits a center-localized magnetization profile, which suggests the correlation between localization and $E_{sp}$ enhancement, independent of the localization site. Additional details on doping-induced effects on TiO$_{2}$ NRs with alternative widths are provided in the supplementary material (see Figs.~\ref{fig:sup_TiO1TUPZNR6} and~\ref{fig:sup_TiO1TUPZNR9}). 

To analyze this correlation in a quantitative manner, the spatial localization of the magnetization profile \(\{m_i\}\) along the width of a hole-doped NR can be quantified employing the inverse participation ratio (IPR), defined as:
\[
\mathrm{IPR}
\;=\;
\frac{\displaystyle\sum_{i} \lvert m_i\rvert^2}
     {\Bigl(\sum_{i} \lvert m_i\rvert \Bigr)^2}\,.
\tag{1}
\]
Consequently, for a magnetization profile described by \(M\) grid points, \(\mathrm{IPR}\) approaches \(1/M\) for a completely delocalized (uniform) distribution \(m_i = 1/M\), whereas it reaches unity if the magnetization is entirely concentrated on a single site (\(m_j=1,\,m_{i\neq j}=0\))~\cite{bell1970atomic}. Hence, larger values of \(\mathrm{IPR}\) correspond to a more strongly localized magnetization profile across the NR width.

\begin{figure}[h!]  
  \centering  
  \includegraphics[width=0.55\linewidth]{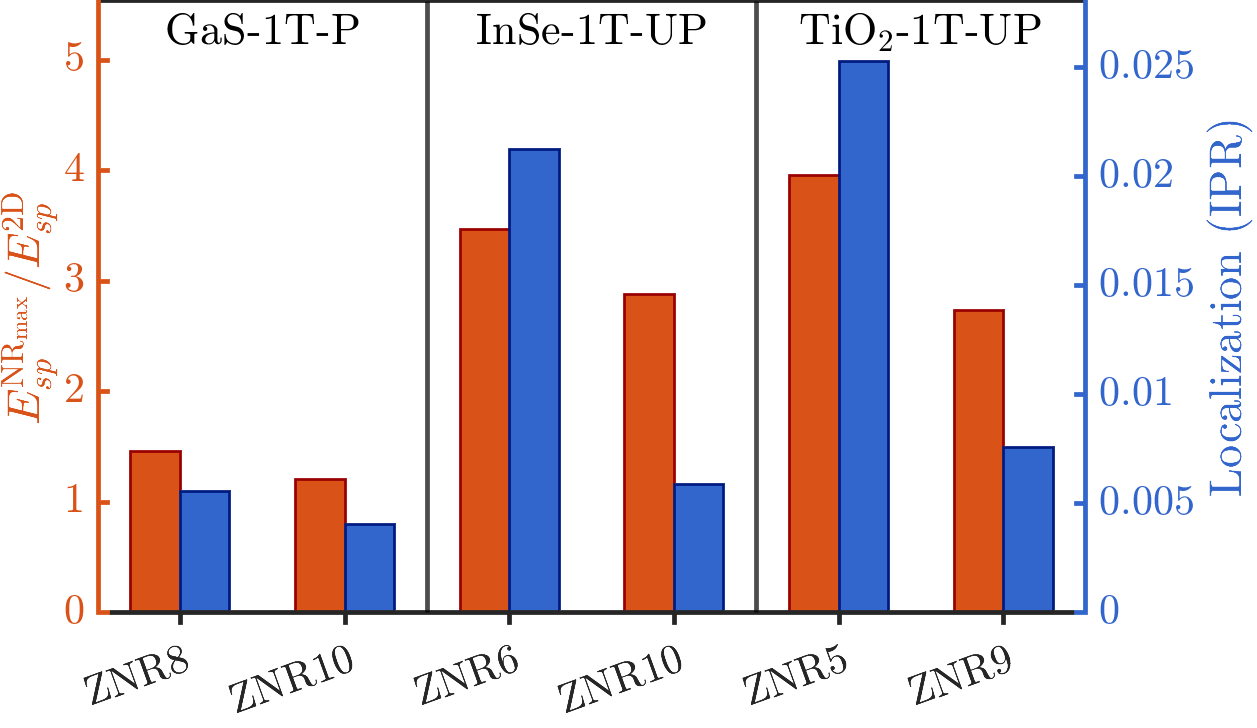}
  \caption{Bar plot of the ratio of highest NR $E_{sp}$ to corresponding 2D $E_{sp}$ at the equivalent carrier density (left axis, orange) and localization quantified by inverse participation ratio (IPR, right axis, blue) of magnetization profile at carrier density corresponding to $E^{NR_{max}}_{sp}$ for selected GaS, InSe, and TiO$_{2}$ ZNRs.}
  \label{fig:local_max_spe}
\end{figure}

In Fig.~\ref{fig:local_max_spe} we present the ratio of highest NR $E_{sp}$ to corresponding 2D $E_{sp}$ at the equivalent carrier density, and localization, quantified by IPR of magnetization profile at carrier density corresponding to $E^{NR_{max}}_{sp}$, for selected GaS, InSe, and TiO$_{2}$ ZNRs. The highest $E_{sp}$ enhancement among the considered structures was observed in InSe and TiO$_{2}$ 1T-UP-ZNRs, which exhibited pronounced localizations in magnetization profiles at the edge and center of NRs, respectively. Nevertheless, from Fig.~\ref{fig:local_max_spe} it can be noticed that InSe 1T-UP-ZNR6 and TiO$_{2}$ 1T-UP-ZNR5 possess roughly the same IPR magnitudes. As IPR is independent of the spatial position of localization, this observation confirms that the correlation between localization and $E_{sp}$ enhancement is independent of the localization site. Furthermore, one can clearly see the consequence of increasing size and appearance of magnetization not only at the edges but also at the center in wider 1T-UP-ZNRs, as shown in the example of InSe 1T-UP-ZNR10 (see Fig.~\ref{fig:sup_InSe1TUPZNR10}), which demonstrates a lower IPR value and SPE relative to the narrower counterpart. Similarly, with increasing width of TiO$_{2}$ 1T-UP-ZNR5 to ZNR9, there is a decrease in both IPR and SPE enhancement magnitudes, as magnetization appears not only at the center but also at the edges of NR (see Fig.~\ref{fig:sup_TiO1TUPZNR9}), leading to lower localization. In addition, the much lower SPE enhancement of GaS 1T-P-ZNRs, compared to 1T-UP-ZNRs, is consistent with lower IPR magnitudes, due to the significant decrease in magnetization localization upon passivation.

\section*{CONCLUSION} 

In quest of enhanced magnetism, we investigated hole-doped nanoribbons derived from 2D materials with Mexican-hat-shaped valence band edges. Our primary motivation was to investigate the potential enhancement resulting from quantum confinement going from 2D to 1D.  We studied various widths of GaS, InSe, and TiO$_{2}$ nanoribbons. At the same doping level, we observed up to 50~\% enhancement in spin-polarization energy ($E_{sp}$) of hydrogen-passivated GaS nanoribbons, which have minimum localization of spin polarization among the three studied materials. The InSe nanoribbons prefer an unpassivated ground state, and hence they have higher localization at the edges. $E_{sp}$ of InSe nanoribbons reaches up to 3.5 times that of its 2D counterpart. Both GaS and InSe belong to the same material family and have similar $E_{sp}$ in 2D. This indicates that $E_{sp}$ is enhanced due to both quantum confinement (from 2D to 1D) and localization. The $E_{sp}$ of TiO$_{2}$ nanoribbons reaches up to nearly 4 times that of its 2D counterpart with a similar strength of localization (IPR) to the InSe case but concentrated towards the center of the nanoribbons. This implies that it is the strength of localization, rather than the location, that determines the enhancement in $E_{sp}$. The TiO$_{2}$ example also shows that similar enhancements can be expected in other material families.

\section*{Acknowledgments}
E.A., A.M. and S.J. acknowledge support from the TÜBİTAK project number 124F108. H.S. acknowledges support from the Air Force Office of Scientific Research (AFOSR, Award No. FA9550-21-1-0261). The calculations were performed at TÜBİTAK ULAKBİM, High Performance and Grid Computing Center (TRUBA resources) and National Center for High-Performance Computing of Turkey (UHeM) under grant number 1007742020.

\bibliography{main}

\renewcommand{\thefigure}{S\arabic{figure}}
\setcounter{figure}{0}
\newcounter{suppfigure}  
\renewcommand{\thefigure}{S\arabic{suppfigure}}

\renewcommand{\thefigure}{S\arabic{figure}} 
\setcounter{suppfigure}{0} 

\renewcommand{\thetable}{S\arabic{table}}
\setcounter{table}{0}
\newpage
\section*{Supplementary Material}

\begin{figure}[h!]  
  \centering  
  \includegraphics[width=0.75\textwidth]{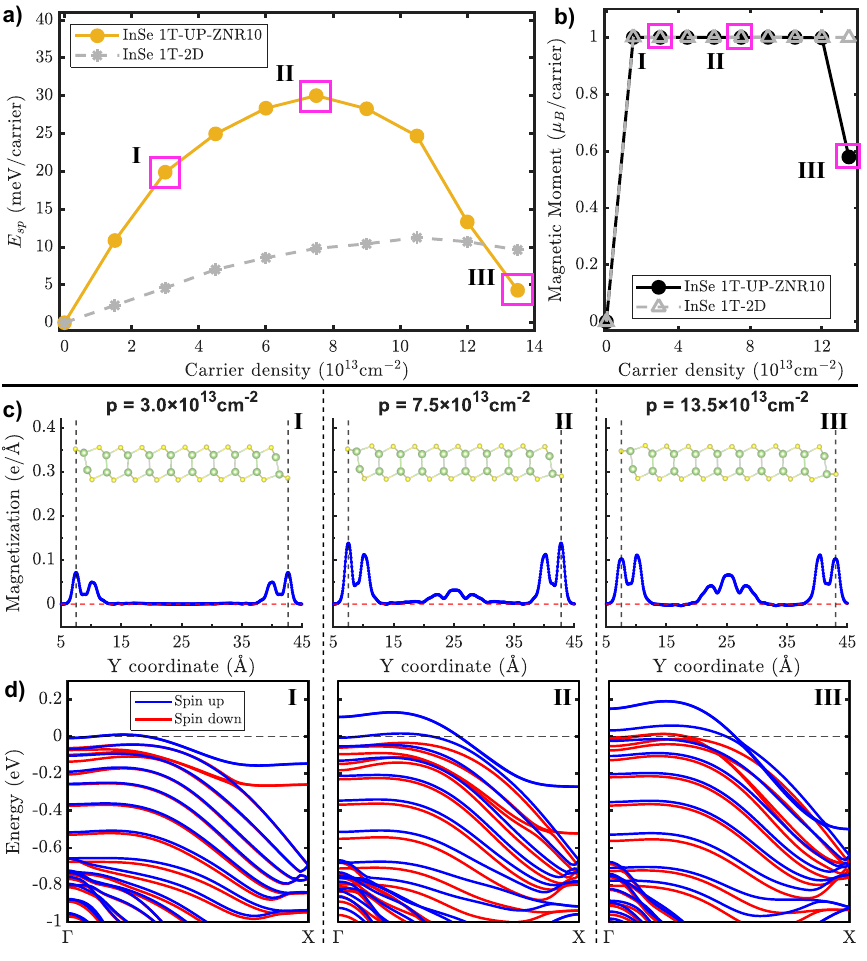}
  \caption{InSe 1T-UP-ZNR10 and 1T-2D a) spin polarization energies and b) magnetic moments per carrier as a function of the carrier density. InSe 1T-UP-ZNR10 c) magnetization profiles along the width of NR and d) spin-polarized band structures at selected carrier densities.}
  \label{fig:sup_InSe1TUPZNR10}
\end{figure}

\begin{figure}[h!]  
  \centering  
  \includegraphics[width=0.55\linewidth]{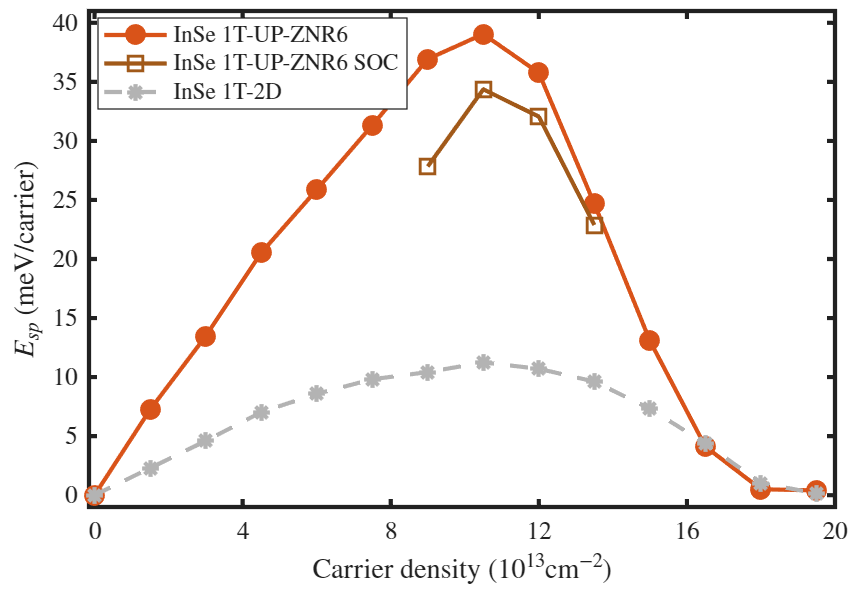}
  \caption{Comparison of spin polarization energies in 1T-phase InSe ZNRs with and without spin–orbit coupling.}
  \label{fig:sup_INSESOC}
\end{figure}

\begin{figure}[h!]  
  \centering  
  \includegraphics[width=0.75\textwidth]{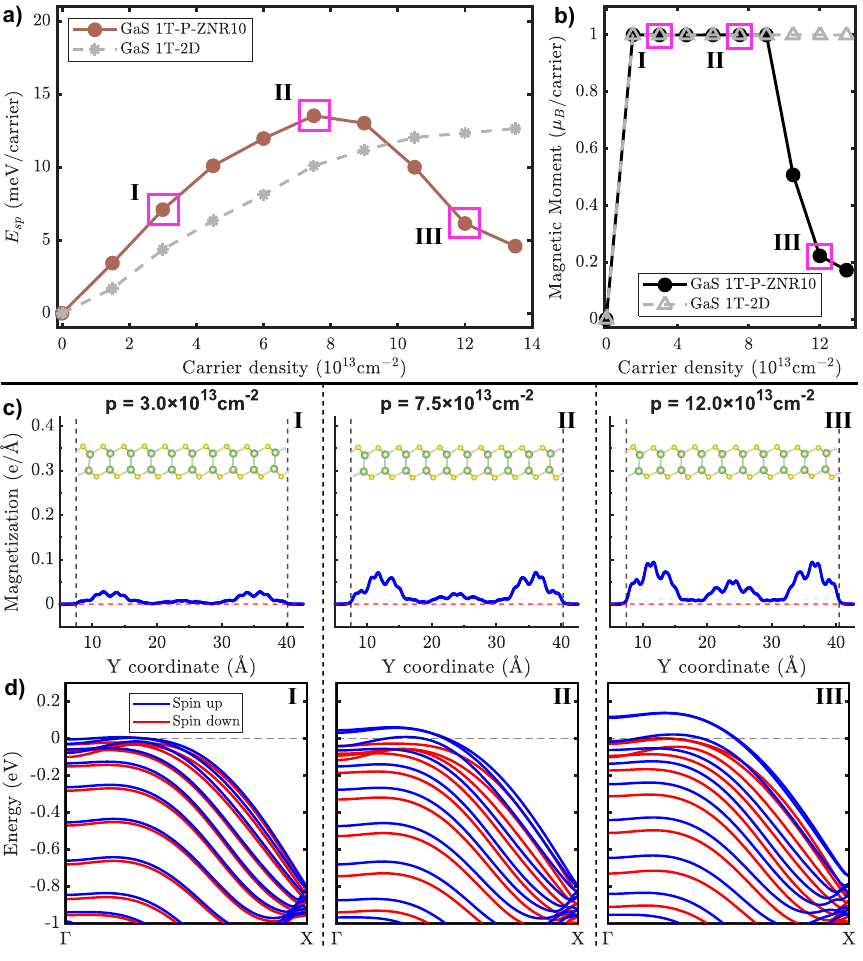}
  \caption{GaS 1T-P-ZNR10 and 1T-2D a) spin polarization energies and b) magnetic moments per carrier as a function of the carrier density. GaS 1T-P-ZNR10 c) magnetization profiles along the width of NR and d) spin-polarized band structures at selected carrier densities.}
  \label{fig:sup_GaS1TPZNR10}
\end{figure}

\begin{figure}[h!]  
  \centering  
  \includegraphics[width=0.75\textwidth]{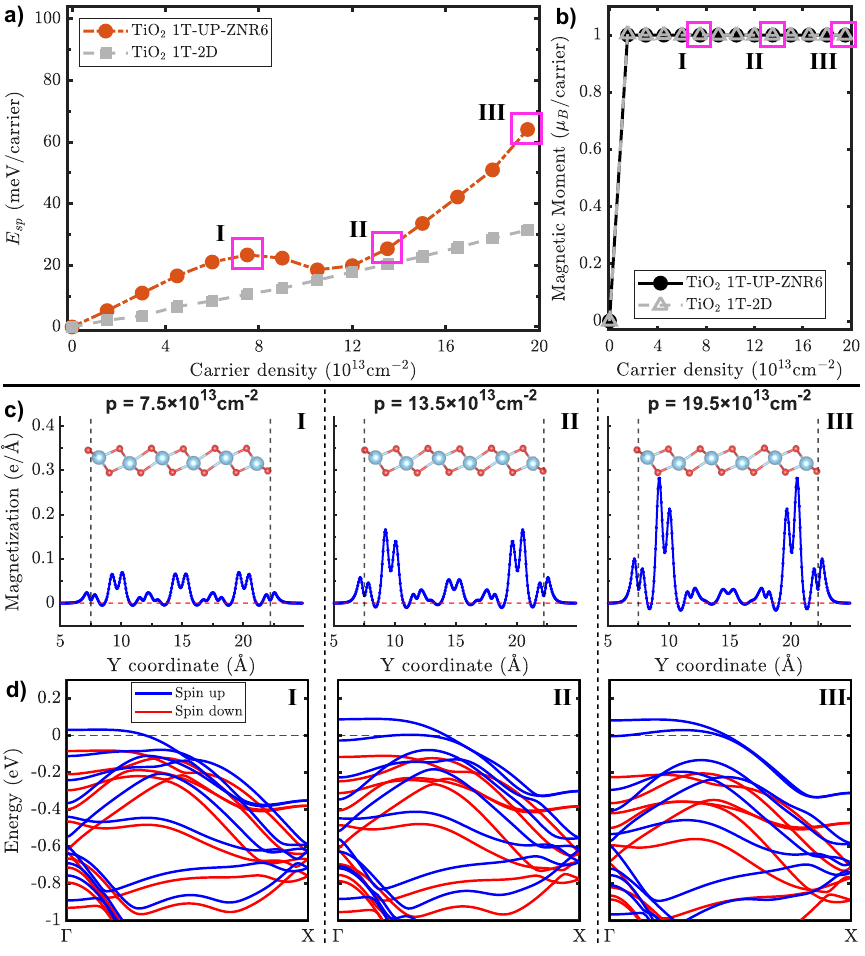}
  \caption{TiO$_{2}$ 1T-UP-ZNR6 and 1T-2D a) spin polarization energies and b) magnetic moments per carrier as a function of the carrier density. TiO$_{2}$ 1T-UP-ZNR6 c) magnetization profiles along the width of NR and d) spin-polarized band structures at selected carrier densities.}
  \label{fig:sup_TiO1TUPZNR6}
\end{figure}

\begin{figure}[h!]  
  \centering  
  \includegraphics[width=0.75\textwidth]{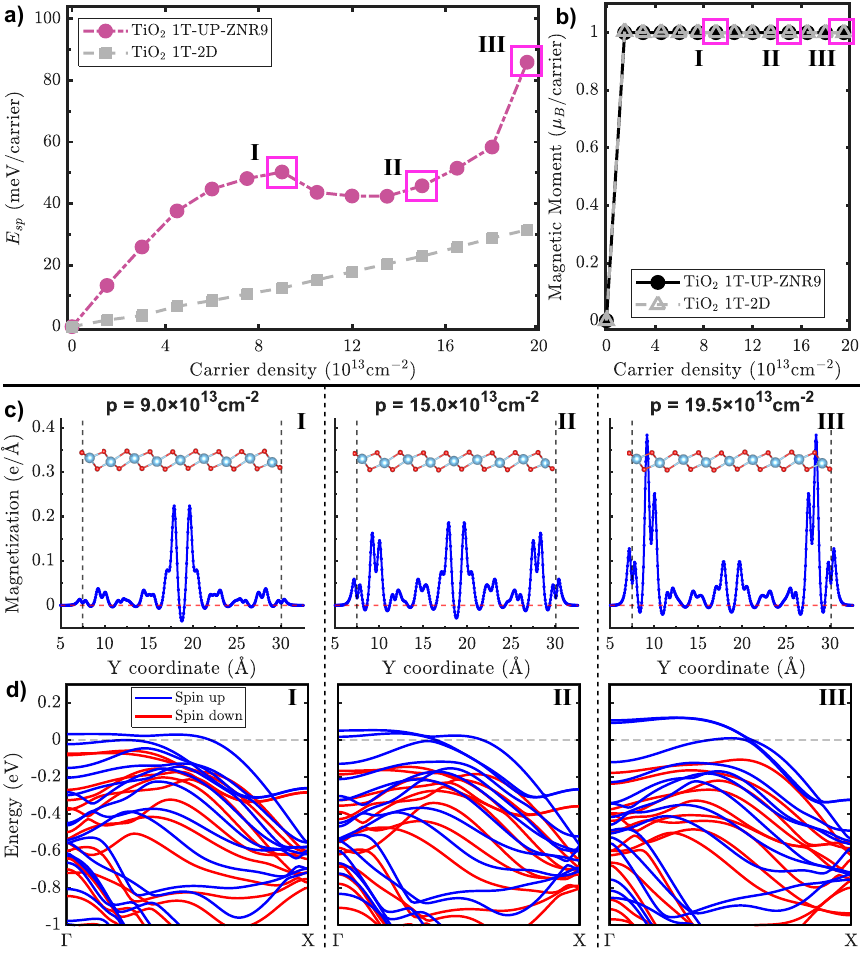}
  \caption{TiO$_{2}$ 1T-UP-ZNR9 and 1T-2D a) spin polarization energies and b) magnetic moments per carrier as a function of the carrier density. TiO$_{2}$ 1T-UP-ZNR9 c) magnetization profiles along the width of NR and d) spin-polarized band structures at selected carrier densities.}
  \label{fig:sup_TiO1TUPZNR9}
\end{figure}

\end{document}